\documentclass[12pt,a4paper]{article}
\pdfoutput=1
\usepackage[utf8]{inputenc}
\usepackage{amsmath,graphicx,dcolumn}
\usepackage{siunitx}
\usepackage[mathscr]{eucal}
\usepackage{cite}
\usepackage{color}
\usepackage{rotating}
\newcommand\as{\alpha_{\mathrm{S}}}

\def\beq{\begin{equation}} 
\def\eeq{\end{equation}} 

\def\to{\rightarrow}

\def\mt{m_t}
\def\rcut{r_{\rm cut}}

\begin{document} 
\begin{titlepage}
\begin{flushright}
ZU-TH 02/19\\
\end{flushright}

\renewcommand{\thefootnote}{\fnsymbol{footnote}}
\vspace*{2cm}

\begin{center}
{\Large \bf Top-quark pair hadroproduction\\[0.2cm] at next-to-next-to-leading order in QCD}
\end{center}

\par \vspace{2mm}
\begin{center}
  {\bf Stefano Catani${}^{(a)}$, Simone Devoto${}^{(b)}$, Massimiliano Grazzini${}^{(b)}$,\\[0.2cm]
Stefan Kallweit${}^{(c)}$,
Javier Mazzitelli${}^{(b)}$}
and
{\bf Hayk Sargsyan${}^{(b)}$}

\vspace{5mm}

${}^{(a)}$INFN, Sezione di Firenze and
Dipartimento di Fisica e Astronomia,\\[0.1cm] 
Universit\`a di Firenze,
I-50019 Sesto Fiorentino, Florence, Italy\\[0.25cm]

${}^{(b)}$Physik Institut, Universit\"at Z\"urich, CH-8057 Z\"urich, Switzerland\\[0.25cm]

$^{(c)}$Universit\`{a} degli Studi di Milano-Bicocca, 20126, Milan, Italy

\vspace{5mm}

\end{center}

\par \vspace{2mm}
\begin{center} {\large \bf Abstract} 

\end{center}
\begin{quote}
\pretolerance 10000

We report on a new calculation of the next-to-next-to-leading order (NNLO) QCD radiative corrections to the inclusive production of top-quark pairs at hadron colliders.
The calculation is performed by using the $q_T$ subtraction formalism to handle and cancel infrared singular contributions at intermediate stages of the computation.
We present numerical results for the total cross section in $pp$ collisions at $\sqrt{s}=8$~TeV and $13$~TeV, and we compare them with those obtained by using the publicly available numerical program {\sc Top++}.
Our computation represents the first complete application of the $q_T$ subtraction formalism to the hadroproduction of a colourful high-mass system at NNLO.

\end{quote}

\vspace*{\fill}
\begin{flushleft}
January 2019
\end{flushleft}
\end{titlepage}

\setcounter{footnote}{1}
\renewcommand{\thefootnote}{\fnsymbol{footnote}}

The top quark ($t$) is the heaviest known elementary particle, and due to its large coupling to the Higgs boson it is expected to play a special role in electroweak symmetry breaking. Studies of top-quark production and decay are central in the LHC physics programme, allowing us to precisely test the Standard Model and, at the same time, offering a window on possible physics beyond the Standard Model. The LHC supplies a huge number of top-quark events, thereby offering an excellent environment for such studies.

Within the Standard Model
the main source of top quarks in hadronic collisions is top-quark pair production. 
Studying the production of $t{\bar t}$ pairs at hadron colliders
can not only shed light on the nature of the electroweak-symmetry breaking, but
it also provides information on the backgrounds of many new-physics models.

Next-to-leading order (NLO) QCD corrections 
to the total cross section for this production process have been computed thirty years ago
\cite{Nason:1987xz, Beenakker:1988bq, Beenakker:1990maa, Nason:1989zy}.
The calculation of the next-to-next-to-leading order (NNLO) QCD 
corrections to the $t{\bar t}$ total cross section was completed a few years ago
\cite{Baernreuther:2012ws, Czakon:2012zr, Czakon:2012pz, Czakon:2013goa}.
Besides the total cross section,  differential cross sections and more general
kinematical distributions are of 
great importance for precision studies \cite{Mangano:1991jk}. 
The $t{\bar t}$ asymmetry, which is non-vanishing starting from the NLO level,
is known up to NNLO \cite{Czakon:2014xsa}.
Other NNLO results on differential distributions 
are available \cite{Czakon:2015owf,Czakon:2016ckf,Czakon:2017dip}.

At the partonic level, the NNLO calculation of $t{\bar t}$ production requires 
the evaluation of tree-level contributions with two additional
partons in the final state, 
of one-loop contributions with one additional parton and 
of purely virtual contributions.
The required tree-level and one-loop scattering amplitudes
are known and they are the same as those entering the NLO calculation of 
the associated production of a $t{\bar t}$ pair and one jet  \cite{Dittmaier:2007wz,Dittmaier:2008uj}.
The purely virtual contributions depend on the two-loop scattering amplitudes
and on the square of one-loop scattering amplitudes.
Partial results for the two-loop amplitude are available in analytic
form \cite{Bonciani:2008az,Bonciani:2009nb,Bonciani:2010mn,Bonciani:2013ywa}, and the complete computation 
has been carried out numerically \cite{Czakon:2008zk,Baernreuther:2013caa}.
The square of the one-loop amplitudes is also known \cite{Korner:2008bn,Anastasiou:2008vd,Kniehl:2008fd}.

The implementation of the above contributions in a (fully differential)
NNLO calculation is a highly non-trivial task because of the presence of infrared (IR) 
divergences at intermediate stages of the calculation. 
Various methods have been proposed and used to overcome these difficulties 
at the NNLO level (see e.g. Ref.~\cite{Bendavid:2018nar} and references therein).

Partial results for $t{\bar t}$ production were obtained by using the antenna subtraction method \cite{GehrmannDeRidder:2005cm, Abelof:2011jv}, 
by considering the $q{\bar q}$ channel
at leading colour and including the light-quark contributions  \cite{Abelof:2014fza, Abelof:2014jna,Abelof:2015lna}.
The only complete NNLO computation for $t{\bar t}$ production to date is that of Refs.~\cite{Baernreuther:2012ws, Czakon:2012zr, Czakon:2012pz, Czakon:2013goa,Czakon:2014xsa,Czakon:2015owf,Czakon:2016ckf,Czakon:2017dip}, which was performed by using the {\sc Stripper} method \cite{Czakon:2010td,Czakon:2011ve,Czakon:2014oma}.

In this Letter we report on a new complete computation of $t{\bar t}$ production at NNLO based on the $q_T$ subtraction formalism~\cite{Catani:2007vq}.
The $q_T$ subtraction formalism is
a method to handle and cancel the IR divergences in QCD computations at NLO and NNLO accuracy.
The method uses IR subtraction counterterms that are constructed by considering
the transverse-momentum ($q_T$) distribution
of the produced high-mass system in the limit $q_T \to 0$.
If the produced high-mass system is composed of non-QCD (colourless) partons
(e.g.\ leptons, vector bosons or Higgs bosons), the behaviour of the 
$q_T$ distribution in the limit $q_T \to 0$ has a universal 
(process-independent) structure that is explicitly known up to NNLO
through the formalism of transverse-momentum resummation 
\cite{Catani:2013tia}. These results on transverse-momentum resummation
are sufficient to fully specify the $q_T$ subtraction formalism for this entire
class of processes. 
In the case of heavy-quark production the transverse-momentum resummation formalism has been developed only recently \cite{Zhu:2012ts,Li:2013mia,Catani:2014qha}.
Nonetheless, such information was already sufficient to apply the $q_T$ subtraction formalism to $t{\bar t}$ production and to obtain
the complete NLO corrections and the NNLO contributions in all the flavour off-diagonal channels \cite{Bonciani:2015sha}.
The NNLO computation in the flavour diagonal channels requires additional perturbative information (see below), and the ensuing results are presented here for the first time.

According to the $q_T$ subtraction method~\cite{Catani:2007vq}, the NNLO
differential cross section $d{\sigma}^{t{\bar t}}_{\rm NNLO}$ for 
the inclusive production 
process $pp\to t{\bar t}+X$
can be written as
\begin{equation}
\label{eq:main}
d{\sigma}^{t{\bar t}}_{\rm NNLO}={\cal H}^{t{\bar t}}_{\rm NNLO}\otimes d{\sigma}^{t{\bar t}}_{\rm LO}
+\left[ d{\sigma}^{t{\bar t}+\rm{jet}}_{\rm NLO}-
d{\sigma}^{t{\bar t}, \, CT}_{\rm NNLO}\right],
\end{equation}
where $d{\sigma}^{t{\bar t}+\rm{jet}}_{\rm NLO}$ is the $t{\bar t}$+jet cross 
section 
at NLO accuracy.
The square bracket term of Eq.~(\ref{eq:main}) is IR finite in the limit
$q_T \to 0$, but its individual contributions,
$d{\sigma}^{t{\bar t}+\rm{jet}}_{\rm NLO}$ and
$d{\sigma}^{t{\bar t}, \, CT}_{\rm NNLO}$, are separately divergent.
The contribution $d{\sigma}^{t{\bar t}+\rm{jet}}_{\rm NLO}$ can be evaluated with any available NLO method to handle and cancel IR divergences.
The IR subtraction counterterm $d{\sigma}^{t{\bar t}, \,CT}_{\rm NNLO}$
is obtained from the NNLO perturbative expansion 
(see e.g.\ Refs.~\cite{Bozzi:2005wk,Bozzi:2007pn,Bonciani:2015sha})
of the resummation formula
of the logarithmically-enhanced
contributions to the $q_T$ distribution
of the $t{\bar t}$ pair \cite{Zhu:2012ts,Li:2013mia,Catani:2014qha}:
the explicit form of $d{\sigma}^{t{\bar t}, \,CT}_{\rm NNLO}$ is fully known.

To complete the NNLO calculation, the second-order functions
${\cal H}^{t{\bar t}}_{\rm NNLO}$ in Eq.~(\ref{eq:main}) are needed.
These functions embody
process-independent and process-dependent contributions. The process-independent contributions to 
${\cal H}^{t{\bar t}}_{\rm NNLO}$ are analogous to those entering Higgs
boson \cite{Catani:2007vq} and vector boson \cite{Catani:2009sm} production,
and they are explicitly known 
\cite{Catani:2011kr,Catani:2012qa,Gehrmann:2012ze,Gehrmann:2014yya}. In the flavour off-diagonal channels 
the process-dependent contributions to ${\cal H}^{t{\bar t}}_{\rm NNLO}$
originate from the knowledge of the one-loop virtual amplitudes of the partonic
processes $q{\bar q} \to t{\bar t}$ and $gg\to t{\bar t}$,
and from the explicit results on the NLO {\em azimuthal correlation} terms
in the transverse-momentum resummation formalism \cite{Catani:2014qha}.
The computation of ${\cal H}^{t{\bar t}}_{\rm NNLO}$ in the {\it diagonal} $q{\bar q}$ and $gg$ channels additionally requires the two-loop amplitudes
for $q{\bar q}\to t{\bar t}$ and $gg\to t{\bar t}$ and the evaluation of new contributions of purely {\it soft} origin.
The two-loop amplitudes  are available in numerical form \cite{Baernreuther:2013caa}, whereas the computation of the additional soft contributions
has been completed by some of us \cite{inprep}.\footnote{An independent computation of these soft contributions has recently been presented in Ref.~\cite{Angeles-Martinez:2018mqh}.}
Therefore we are now in a position to complete the calculation of
Ref.~\cite{Bonciani:2015sha} and to obtain the full NNLO cross section.

Before presenting our results we briefly describe our implementation.
The NNLO cross section can be expressed as
\begin{equation}
  \sigma_{\rm NNLO}=\sigma_{\rm NLO}+\Delta\sigma_{\rm NNLO}\, .
\end{equation}
The NLO contribution $\sigma_{\rm NLO}$ is evaluated by using the
{\sc Munich} code \cite{Kallweit:Munich}, which
provides a fully automated implementation of the 
NLO dipole subtraction formalism \cite{Catani:1996jh,Catani:1996vz,Catani:2002hc}.
We use Eq.~(\ref{eq:main}) to compute the NNLO correction $\Delta\sigma_{\rm NNLO}$.
The NLO cross section $d{\sigma}^{t{\bar t}+\rm{jet}}_{\rm NLO}$ is computed by using {\sc Munich}.
The subtraction counterterm $d{\sigma}^{t{\bar t}, \, CT}_{\rm NNLO}$ is also implemented in {\sc Munich}, whereas the contribution proportional to ${\cal H}^{t{\bar t}}_{\rm NNLO}$ is evaluated with an
extension of the numerical programs developed for Higgs boson \cite{Catani:2007vq} and vector-boson \cite{Catani:2009sm} production.
All the required (spin- and colour-correlated) tree-level and one-loop amplitudes are obtained by using {\sc OpenLoops} \cite{Cascioli:2011va}, except for the four-parton tree-level colour correlations that we obtain through an analytic implementation.
{\sc OpenLoops} relies on the fast and stable tensor reduction of {\sc Collier} \cite{Denner:2014gla,Denner:2016kdg}, supported by a rescue system based on quad-precision {\sc CutTools} \cite{Ossola:2007ax} with {\sc OneLOop} \cite{vanHameren:2010cp} to deal with exceptional phase-space points.
To the purpose of validating our results for the real--virtual contribution,
we have used also the new on-the-fly reduction of {\sc OpenLoops\,2}~\cite{Buccioni:2017yxi, openloops2} and the independent matrix-element generator {\sc Recola}~\cite{Denner:2017wsf}, finding complete agreement.

The contribution in the square bracket in Eq.~(\ref{eq:main}) is formally finite in the limit $q_T \to 0$, but both $d{\sigma}^{t{\bar t}+\rm{jet}}_{\rm NLO}$ and
$d{\sigma}^{t{\bar t}, \, CT}_{\rm NNLO}$ are separately divergent. In practice, the computation is carried out by introducing a cut-off $\rcut$ on the dimensionless variable $r=q_T/M$, where $M$ is the invariant mass of the $t{\bar t}$ pair. The final result is obtained by performing the limit $\rcut\to 0$. To do so, the cross section is computed at fixed values of $\rcut$ in the interval $[0.01\%, r_{\rm max}]$. Quadratic least $\chi^2$ fits are performed by varying $r_{\rm max}$ from $0.5\%$ to $1\%$. The result with the lowest
$\chi^2/$degrees-of-freedom value is kept as the best fit. The extrapolation uncertainty is determined
by comparing the result of the best fit with the results of the other fits.
This procedure is the same as implemented in {\sc Matrix} \cite{Grazzini:2017mhc}, and it has been shown to provide a conservative estimate of the systematic uncertainty in the $q_T$ subtraction procedure for various processes (see Sec.~7 in Ref.~\cite{Grazzini:2017mhc}).

To present our quantitative results, we consider $pp$ collisions at $\sqrt{s}=8$~TeV and 13 TeV, and
we use the NNPDF31 \cite{Ball:2017nwa} NNLO parton distribution functions throughout. The QCD running of $\as$ is evaluated at three-loop order with $\as(m_Z)=0.118$, and
the pole mass of the top quark is fixed to $\mt=173.3$ GeV. The central values of the renormalization ($\mu_R$) and factorization ($\mu_F$) scales are fixed to $\mu_R=\mu_F=\mt$.
{\renewcommand{\arraystretch}{1.6}
\begin{table}
\begin{center}
\begin{tabular}{|c|c|c|}
\hline
$\sigma_\text{NNLO}$ [pb] & $q_T$ subtraction & {\sc Top++} \\ \hline
8 TeV  & $238.5(2)^{+3.9\%}_{-6.3\%}$ & $238.6^{+4.0\%}_{-6.3\%}$ \\
13 TeV  & $793.4(6)^{+3.5\%}_{-5.7\%}$ & $794.0^{+3.5\%}_{-5.7\%}$ \\ \hline
\end{tabular}
\end{center}
\caption{
Total cross section for $t\bar{t}$ production in $pp$ collisions. The quoted uncertainties are obtained through scale variations as described in the text. Numerical uncertainties on the last digit are stated in brackets (and include the $\rcut\to 0$ extrapolation uncertainties).
}
\label{table:totalXS}
\end{table}
We start the presentation by considering the complete NNLO cross sections. In Table~\ref{table:totalXS} our results at $\sqrt{s}=8$~TeV and 13 TeV are compared with the corresponding results obtained with the numerical program {\sc Top++} \cite{Czakon:2011xx}\footnote{The program {\sc Top++} is used with the input parameter {\tt Precision=3}.},
which implements the NNLO calculation of
Refs.~\cite{Baernreuther:2012ws, Czakon:2012zr, Czakon:2012pz, Czakon:2013goa}
(at NLO {\sc Top++} uses the parametrization of Refs.~\cite{Langenfeld:2009wd,Aliev:2010zk} of the analytic result of Ref.~\cite{Czakon:2008ii}). In Table~\ref{table:totalXS} the NNLO cross sections are reported with their perturbative uncertainty, which is estimated through the customary procedure of independently varying $\mu_R$ and $\mu_F$ by a factor of two around their central value with the constraint $0.5\leq \mu_F/\mu_R\leq 2$.
The program {\sc Top++} gives results without an associated numerical error.
Our results are given with an uncertainty that is obtained by combining statistical errors from the Monte Carlo integration and the systematic uncertainty associated to the $\rcut \to 0$ extrapolation. Such combined uncertainty turns out to be at the {\it per mille} level, and our results are consistent with those of {\sc Top++} for all the considered values of $\mu_R$ and $\mu_F$.

In  Table \ref{table:channels} the NNLO corrections $\Delta\sigma_{\rm NNLO}$ in the various partonic channels $ab\to t{\bar t}+X$ computed with $q_T$ subtraction are compared
to the corresponding results obtained with {\sc Top++}.
The contribution from all the channels
with $ab= qg,{\bar q}g$ is labelled as $qg$, and the 
contribution from all the channels with 
$ab=qq, {\bar q}{\bar q}, qq', {\bar q}{\bar q}', q{\bar q}', {\bar q} q'$
is labelled as $q({\bar q})q^\prime$.

{\renewcommand{\arraystretch}{1.6}
\begin{table}
\begin{center}
\begin{tabular}{|c|S[table-format=3.7]|S[table-format=3.7]|S[table-format=3.4]|S[table-format=3.4]|}
\hline
$~$ & \multicolumn{2}{c|}{$q_T$ subtraction} & \multicolumn{2}{c|}{{\sc Top++}} \\ \hline
$\Delta\sigma_\text{NNLO}$ [pb]  & $\text{8 TeV}$ & $\text{13 TeV}$ & $\text{8 TeV}$ & $\text{13 TeV}$ \\ \hline
$gg$  & 25.77(23) & 80.99(54) & 25.86 & 81.54 \\
$q\bar{q}$  & 2.249(12) & 4.713(16) & 2.248 & 4.739 \\
$qg$  & -2.349(31) & -4.16(19) & -2.340 &  -4.089 \\
$q({\bar q})q^\prime$ & 0.1563(11) & 0.6378(34) & 0.1563 & 0.6375 \\ \hline
\end{tabular}
\end{center}
\caption{
NNLO corrections $\Delta\sigma_{\rm NNLO}$, split into the different production channels, for $\mu_R = \mu_F = m_t$. Numerical integration errors on the last digits are stated in brackets.
}
\label{table:channels}
\end{table}

We see that the numerical uncertainties of our NNLO corrections are at the percent level or smaller, except for the $qg$ contribution at $\sqrt{s}=13$ TeV, for which there is a large cancellation between the two terms in Eq.~(\ref{eq:main}) (the term that is proportional to ${\cal H}^{t{\bar t}}_{\rm NNLO}$ and the term in the square bracket). Similar effects were already observed in Ref.~\cite{Bonciani:2015sha}. Comparing our 8\,TeV results for $\Delta\sigma_{\rm NNLO}$ with those obtained by using {\sc Top++}, we see that they are fully compatible within $1\sigma$. At 13\,TeV we also find agreement at the $1\sigma$ level apart from the $q\bar{q}$ channel that exhibits a $1.6\sigma$ difference, which corresponds to about $0.5\%$ of $\Delta\sigma_{\rm NNLO}$ in this channel. Considering the partly statistical nature of our error estimate and the fact that the uncertainties from {\sc Top++} were completely neglected in this discussion, we can state that our results are in agreement with the {\sc Top++} results throughout.

The quality of the $\rcut\to 0$ extrapolation can be assessed by investigating the behaviour of the cross section at fixed values of $\rcut$.
In Fig.~1 we study this behaviour in the different partonic channels. We see that the $\rcut$ dependence is larger than what is observed in the case of the production of a colourless system (see Sec.~7 of Ref.~\cite{Grazzini:2017mhc}), where the powerlike dependence of the total cross section on $\rcut$ is known to be quadratic (modulo logarithmic enhancements). In the case of $t{\bar t}$ production, due to the additional contribution of soft radiation from the heavy quarks, the $\rcut$ dependence of the cross section is expected to be {\it linear} \cite{Catani:2017tuc}, thereby implying a stronger sensitivity to the parameter $\rcut$. The exception is the $q({\bar q})q^\prime$ channel, where the expected $\rcut$ dependence should be quadratic, since this channel does not receive contributions from soft radiation at NNLO.

\begin{figure}[t]
\begin{center}
\begin{tabular}{cc}
  \includegraphics[width=0.48\textwidth]{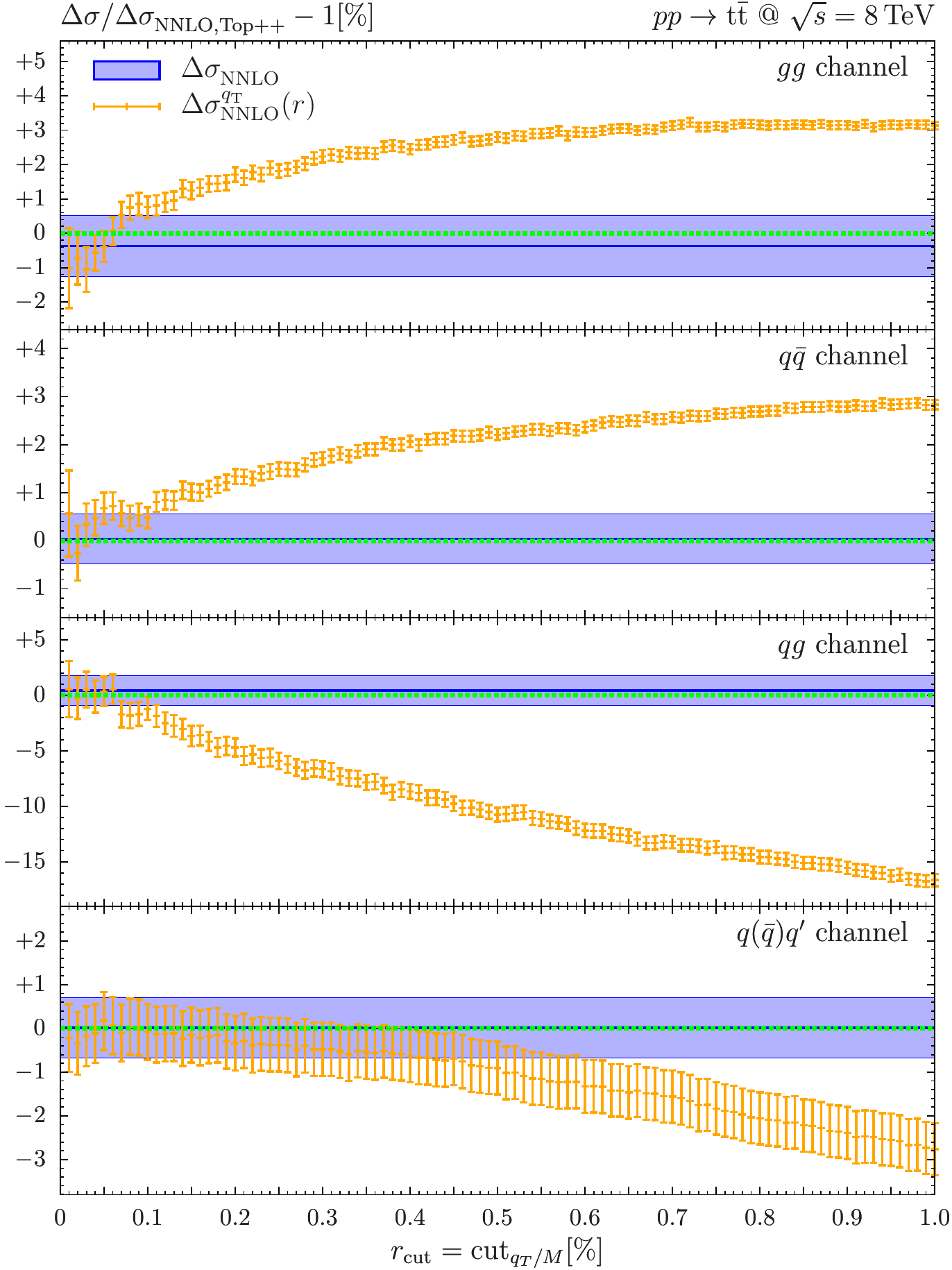} &
  \includegraphics[width=0.48\textwidth]{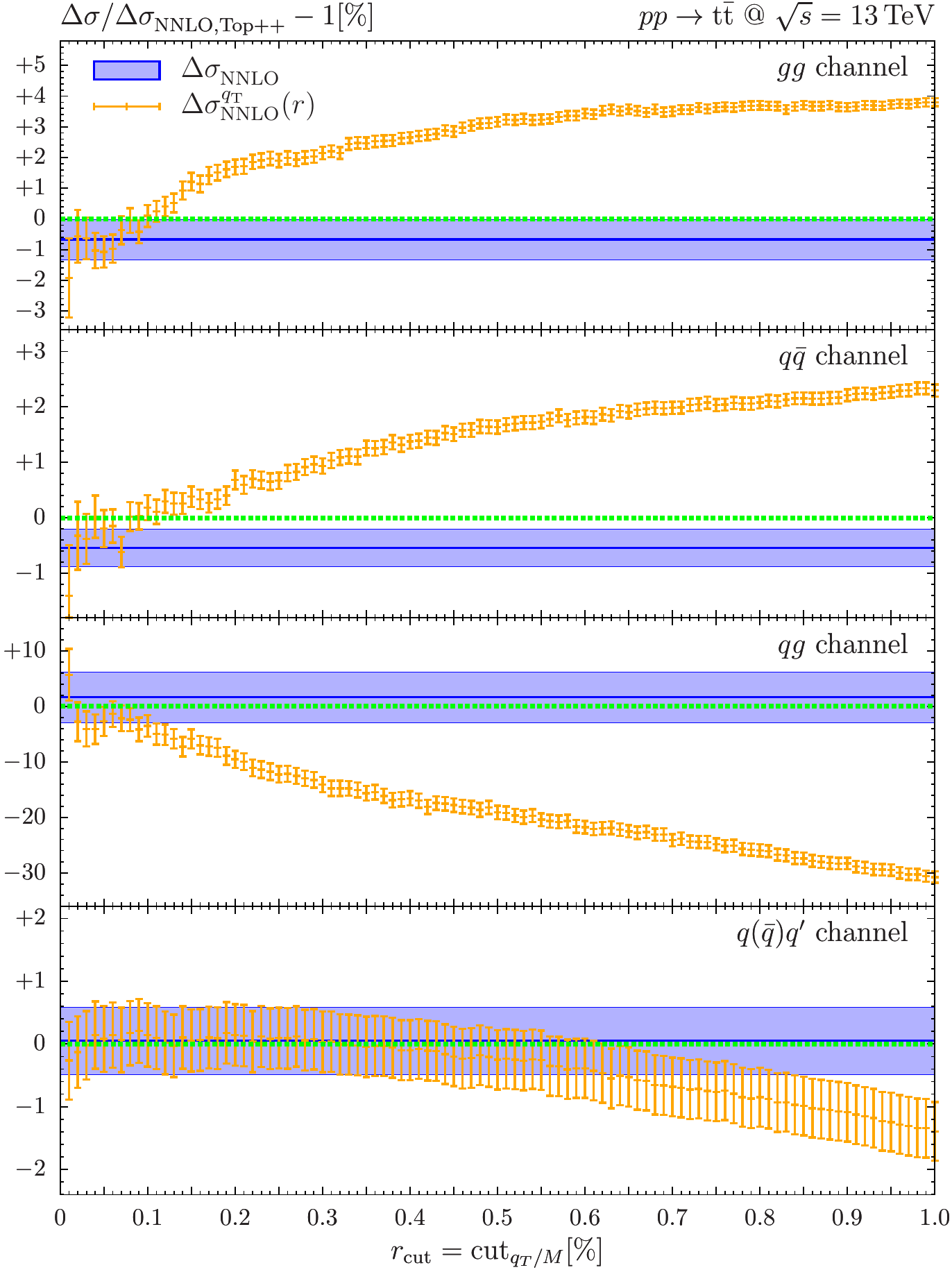}
\end{tabular}                                                                                                                                       
\end{center}
\caption[]{\label{fig:rcut}{NNLO corrections $\Delta\sigma_{\rm NNLO}$ normalized to the {\sc Top++} result as a function of $\rcut$ in the various channels. The blue bands represent our final extrapolated result with its uncertainty.}}
\end{figure}

To conclude, we have reported on a new complete computation of the $t{\bar t}$ cross section in hadron collisions at NNLO in QCD perturbation theory. The computation is performed by
combining tree-level and one-loop QCD amplitudes, as obtained from {\sc OpenLoops}, with two-loop contributions available from the literature. The results are obtained
by using the $q_T$ subtraction formalism to handle and cancel IR singularities. The contributions needed to apply $q_T$ subtraction to this process that were previously unknown
have been computed by some of us, and they will be reported in a separate publication. We have presented
numerical results in $pp$ collisions at 8~TeV and 13~TeV and compared them to the corresponding results obtained with
the  numerical program {\sc Top++}. We find good agreement within the numerical uncertainties. Our computation represents the first complete application of the $q_T$ subtraction formalism to the hadroproduction of a colourful high-mass system at NNLO. The computation can be naturally
extended to differential distributions and by applying arbitrary IR safe cuts on the $t{\bar t}$ pair and the associated QCD radiation. More details on the calculation and additional results will be presented elsewhere.

\noindent {\bf Acknowledgements}. We are most grateful to Federico Buccioni, Jean-Nicolas Lang, Jonas Lindert, Stefano Pozzorini and Max Zoller for their continuous assistance on issues related to {\sc OpenLoops} during the course of this project and for allowing us to use a preliminary version of {\sc OpenLoops\,2}. This work is supported in part by the Swiss National Science Foundation (SNF) under contract 200020-169041. The work of SK is supported by the ERC Starting Grant 714788 REINVENT. 

\pagebreak


\begin{thebibliography}{90}
\bibitem{Kallweit:Munich}
S.~Kallweit,
{\sc Munich} is the abbreviation of ``MUlti-chaNnel Integrator at Swiss~(CH) precision''---an automated parton level NLO 
generator. 
In preparation.
\end{thebibliography}
\end{document}